\documentclass[sigconf]{acmart}

\setcopyright{none}
\renewcommand\footnotetextcopyrightpermission[1]{}

\usepackage[utf8]{inputenc}
\usepackage[english]{babel}

\usepackage{pifont}

\usepackage{xspace}
\newcommand{\eg}{\textit{e.g.,}~}
\newcommand{\ie}{\textit{i.e.,}~}
\newcommand{\cf}{\textit{cf.,}~}

\newcommand{\one}{({\em i})\xspace}
\newcommand{\two}{({\em ii})\xspace}
\newcommand{\three}{({\em iii})\xspace}

\makeatletter
\renewcommand{\paragraph}[1]{\vspace*{0.03in}\noindent{\bf #1.}\hspace{0.25ex \@plus1ex \@minus.2ex}}
\makeatother

\usepackage{booktabs}

\usepackage{balance}
\usepackage{url}
\usepackage{graphicx}
\usepackage{caption}
\usepackage[position=b]{subcaption}

\usepackage{tabularx}
\usepackage{multirow}
\usepackage{float}
\usepackage{makecell}
\usepackage{color, colortbl}

\definecolor{Gray}{gray}{0.9}

\usepackage{pgfplots}
\usetikzlibrary{positioning}
\usepackage{tikz}
\usepackage{tikzpeople} 
\usetikzlibrary{arrows, backgrounds, fit, positioning, quotes, shapes, patterns, hobby}
\usepackage{standalone}
\usepackage{geometry}

\graphicspath{{figs/}}

\begin{document}

\title[A Reproducibility Study of ``IP Spoofing Detection in Inter-Domain Traffic'']{A Reproducibility Study of \\\ ``IP Spoofing Detection in Inter-Domain Traffic''}

\author {Jasper Eumann}
\affiliation{%
  \institution{HAW Hamburg}
}
\email{jasper.eumann@haw-hamburg.de}

\author {Raphael Hiesgen}
\affiliation{%
  \institution{HAW Hamburg}
}
\email{raphael.hiesgen@haw-hamburg.de}

\author {Thomas C. Schmidt}
\affiliation{%
  \institution{HAW Hamburg}
}
\email{t.schmidt@haw-hamburg.de}

\author {Matthias W\"ahlisch}
\affiliation{%
  \institution{Freie Universit\"at Berlin}
}
\email{m.waehlisch@fu-berlin.de}

\begin{abstract}
IP spoofing enables reflection and amplification attacks, which cause major threats to the current Internet infrastructure.
	Detecting IP packets with incorrect source addresses would help to improve the situation. This is easy at the attacker's network, but very challenging at Internet eXchange Points (IXPs) or in transit networks.
In this reproducibility study, we revisit the paper \textit{Detection, Classification, and Analysis of Inter-Domain Traffic with Spoofed Source IP Addresses} published at ACM IMC~2017~\cite{lskrf-dcait-17}.
Using data from a different IXP and from a different time, we were not able to reproduce the results.
Unfortunately, our further analysis reveals structural problems of the state of the art methodology, which are not easy to overcome. 
\end{abstract}



\maketitle

\section{Introduction}
\label{sec:intro}

IP spoofing injects packets that include an IP source address, which is not advertised to the routing by its origin network.
Consequently, any reply is directed not to its origin but to a different destination. Lack of IP source address validation (SAV) has been identified early as a serious security flaw on the Internet~\cite{b-sptps-89}. 
In combination with a distributed amplification, in which small requests trigger much larger replies, this leads to serious denial of service attacks in the current Internet~\cite{jkkrs-mtuam-17,rowrs-adads-15}.

The most effective approach to SAV and to mitigating reflection attacks~\cite{rowrs-adads-15} is ingress filtering at the network of the attacker~\cite{RFC-2827,RFC-3704}.
This solution, however, is not sufficiently deployed~\cite{ffgmr-manrs-18}.
An alternative solution~\cite{lskrf-dcait-17} proposes a heuristic at central Internet eXchange Points (IXPs) following the observation that a valid packet should flow compliant with control plane information, and hence should reach the IXP via a customer cone that contains its origin. This paper recently published  at ACM IMC 2017 claims that a method is presented ``to passively detect packets with spoofed IP addresses [\dots] and minimize false positive inferences''~\cite[\S~1]{lskrf-dcait-17}. Central to this approach is a reliable inference of  (extended) customer cones from BGP data, which poses the major challenge.

In this paper, we report on our attempts to reproduce the current state of the art, based on a different team and setup~\cite{bbfkp-dbgre-19}. 
At  different times, we analyze data from a large regional IXP instead of data from a large European IXP, which should not affect the validity of the method.  We restrict our work to (all steps of) the methodology (\S~3) presented in~\cite{lskrf-dcait-17}, \ie  the purely algorithmic identification of the customer cones, and purposefully omit the steps of manual cone shaping also performed in~\cite{lskrf-dcait-17}. This work extends our initial presentation at IMC'19~\cite{ehsw-rssdi-19s} with a background analysis.
 
Unfortunately, our findings largely differ from those presented in IMC'17, even though we explore various ways of inferring the customer cones. In particular, spoofed traffic classified in our experiments exceeds the values of IMC'17 by orders of magnitude with a traffic mix that strongly indicates a dominant portion of false positives. We identify plausible reasons for these discrepancies from further analyses and illustrate the underlying structural problems.
It is worth noting that our insights are independent of the vantage point and time but highlight intrinsic drawbacks of the previous methodology.
Independent parallel research~\cite{mlhcb-cisti-19}  confirms parts of our results. We compare and analyze this parallel research, as well.

In the remainder, we recap the methodology in Section~\ref{sec:methodology}, introduce our implementation, the data sources, and discuss the properties of the IXPs under consideration in Section~\ref{sec:implementation}. We present and discuss results of our reproduction attempts in Section~\ref{sec:results}. A deeper discussion in Section~\ref{sec:discussion} identifies conceptual problems of the method presented in IMC'17, and explains the structural reasons for its failures.  Finally, we conclude in Section~\ref{sec:conclusion}.

\section{Recap of IMC'17 Methodology}
\label{sec:methodology}

The objective of the proposed approach~\cite{lskrf-dcait-17} is to sort  \textbf{invalid} (spoofed) traffic from \textbf{regular} (non-spoofed).  Before classifying packets into these two categories the traffic is sanitized by filtering \textbf{bogon} packets, \ie packets with addresses from private networks and other ineligible routable prefixes~\cite{RFC-1918,RFC-5735,RFC-6598}, as well as \textbf{unrouted} packets, \ie packets from sources that do not show any announcements.

For the remaining packets entering the IXP via an IXP member, a check is performed whether each packet arrives via a customer cone that covers the prefix of the origin AS. Such a customer cone includes all ASes that receive (indirect) upstream via the IXP member and includes transitive~peering.  

Due to the limited visibility of BGP relations and the lack of vantage points, it is a major challenge to correctly infer these cones.  Three algorithmic approaches are proposed in~\cite{lskrf-dcait-17} (names taken from the paper):

\paragraph{(1) Naive Approach} Built from public BGP information, this approach considers a packet valid if it originates from an AS that is part of an announced path for its source prefix. It aims to reflect the topology but falls short in representing business relationships between ASes accurately. Live data provides sufficient information to deploy it.

\paragraph{(2) CAIDA Customer Cone} In contrast to the naive cone, CAIDA represents the business relationships rather than the topology. It is created by information such as community strings, directly reported relations, and historic information.  Further details are available in \textit{AS Relationships, Customer Cones, and Validation} by Luckie et al.~\cite{lhdgc-arccv-13}.

\paragraph{(3) Full Cone} This extended cone is built from the assumption that  ASes neighboring  in an announcement are transitively peering. Built from public BGP announcements this approach adds transitive relationships between all peers. Even though this might misinterpret or miss  business relationships, it results in the largest cone and hence reduces the number of invalids.

\paragraph{\textbf{Multi-AS Organization Extension}:} This add-on can be combined with the CAIDA Customer Cone and the Full Cone. It adds information about sibling ASes by building connections between ASes belonging to the same multi-AS organization~\cite{chkw-tam-10}, thus allowing a bidirectional data exchange between them. \medskip

Using these three cones, packets are proposed to be classified either as \textbf{invalid} (spoofed) or as \textbf{regular}.
The full cone approach is the main method examined in the IMC'17 paper and the basis for most of its evaluation. In our reproducibility study, we consider the  various approaches of cone construction and analyse its different impacts on the packet classification.

\section{Implemention and Data Sources}
\label{sec:implementation}

\subsection{Software}
In this this study, we performed replicability and reproducibility work in two phases. First, we replicated the results by applying scripts kindly provided by the IMC'17 authors~\cite{Lichtblau18} to our data sets. These scripts only support constructing the \textit{full cone} from BGP data, which is taken as the indicator of \textit{invalid} packets. We needed to augment these scripts with helper tools for \one reading output from the tool \texttt{bgpdump}, and \two detecting \textit{bogon} and \textit{unrouted} packets. Replicating the results for the full cone served as baseline to verify that our data can be processed as expected. All further results in this paper are based on the second phase, the reproduction.

For the reproduction, we re-implemented all methods to construct all cones (\ie Naive, CAIDA, and Full cone) and to detect bogon and unrouted packets.
Based on this reimplementation, we added enhanced features for classifying payloads of spoofed traffic using \texttt{libpcap}\footnote{\url{https://www.tcpdump.org/}}.
While carefully confirming consistency of the results with the original scripts where applicable, our extended toolset allows  for a faster and more accurate analysis of the classification, as discussed later in more detail.
Our software is publicly available at \url{https://github.com/inetrg/reproducibility-study-ixp-spoofing}.



\subsection{Data Sources}
Our traffic analysis is based on sampled flow data from a large regional Internet Exchange Point in Europe. 
We consider two different time periods, February~19-25, 2018 and June~1-7,~2019. All results shown in this paper are based on the week in  February 2018, while we used the June 2019 data to verify the stability of our results.

To construct the cones and to identify unrouted prefixes, we utilize BGP data from all route collectors available in BGPStream~\cite{okggd-bsflh-16}, for the corresponding weeks as well as one day before and one day~after.

\subsection{Comparing IXPs}
In the two studies under comparison, data from different IXPs have been used. Both IXPs are located in Europe and follow the model of a layer 2 infrastructure that is distributed across several data centers.  Both IXPs sample flow data at comparable frequency (1:10.000 vers. 1:16.000). We focus on one metropolitan area each. The large European IXP has about seven times more actively peering members compared to the large regional counterpart.

When measuring inter-domain data flows between providers and considering their customer cones, characteristics of IXP members and the nature of peering relations is of particular importance. Both IXPs exhibit a heterogeneous member mix, including Tier~1 and large transit ISPs, content providers, national service providers, enterprise networks, as well as educational or research organisations which own ASes. Using information available in PeeringDB~\cite{peeringdb} we can confirm that the relative amount of IXP members per network type is almost equal between both IXPs.
Largest divergences of 10~\% occur for the amount of content networks and Cable/DSL/ISP networks respectively.
It is worth noting that both IXPs operate a route server to facilitate public peering, but private peering relations are also deployed. 

As such, both IXPs cover the peering spectrum well enough to represent  characteristic traffic exchanges and grant comparative insights.

\section{Results}
\label{sec:results}

\begin{table}[b]
  \caption{Comparison of the different classification results for anomalous
  traffic. Significant differences highlighted in~gray.}
  \label{tab::results}
  \begin{tabular}{l l l l l}
    \toprule
    & \multicolumn{2}{c}{IMC 2017} & \multicolumn{2}{c}{This study}
    \\ \cmidrule(r){2-3} \cmidrule(l){4-5}
    & Bytes & Packets & Bytes & Packets
    \\
    \midrule
    Bogon & 0.003\%  & 0.02\% & 0.0009\% & 0.0022\%
    \\
    Unrouted & 0.004\%  & 0.02\%  & 0.00001\% & 0.0001\%
    \\ \midrule
   
    Invalid
    \\
    \quad Naive & 1.1\%   & 1.29\% & 0.579\%  & 1.537\%
    \\ 
   \quad CAIDA & 0.19\%  & 0.3\% & 0.955\% & 1.563\%
    \\
    \quad Full & \cellcolor{Gray}{0.0099\%}  & \cellcolor{Gray}{0.03\%}  & \cellcolor{Gray}{0.2\%}  & 
\cellcolor{Gray}{0.488\%}
    \\
    \bottomrule
  \end{tabular}
\end{table}

We show a first glance of the overall results in Table~~\ref{tab::results}. Our results resemble those of IMC'17 for the amount of invalids based on the naive and CAIDA cones but diverge significantly for invalids based on the full cone and the classification of bogon and unrouted traffic.

It is worth noting that the method of classifying bogon and unrouted traffic is well-known and not specific to the IMC'17 proposal. Observed divergences are not surprising but rather reflect different states of deployment in time and place. For example, as bogon traffic is a typical anomaly from misconfigured NATs, fluctuations between deployments are likely.

The decline of invalids in the full cone, however, does surprise. IMC'17 finds 20 times less invalid bytes after extending the CAIDA cone with additional peering links, whereas we see only a reduction to more than one fifth. Similarly, packet numbers drop for IMC'17 down to one tenth, while ours decrease by a factor of three.
The largely enhanced impact of the IMC'17 full cone leads to a notable reduction of traffic classified as invalid. Neither by adding (fairly effectless) multi-AS organisation extensions (see Section \ref{sec:methodology}), nor by varying input data sets, we could reproduce these strong effects.  

In the remainder of this section, we will analyze the invalid traffic and the impact of cone construction in more detail.

\begin{figure}[t]
  \begin{subfigure}{.45\textwidth}
    \includegraphics[width=\linewidth]{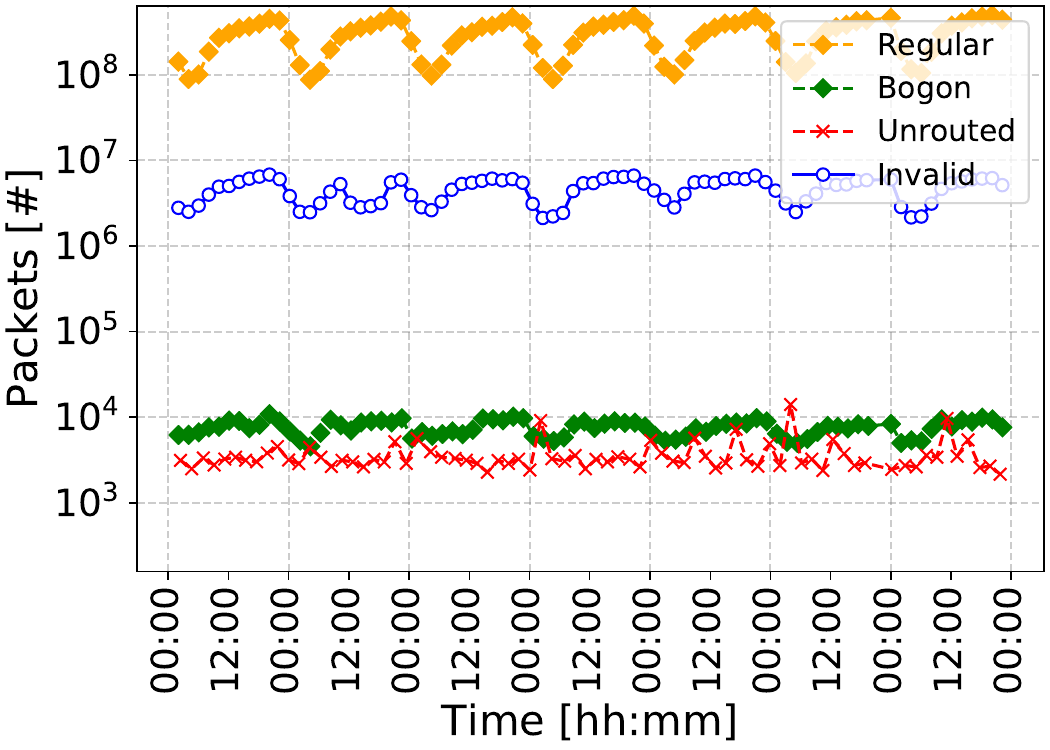}
    \caption{Naive Approach}
    \label{fig::week_naive}
  \end{subfigure}

  \bigskip
  \begin{subfigure}{.45\textwidth}
    \includegraphics[width=\linewidth]{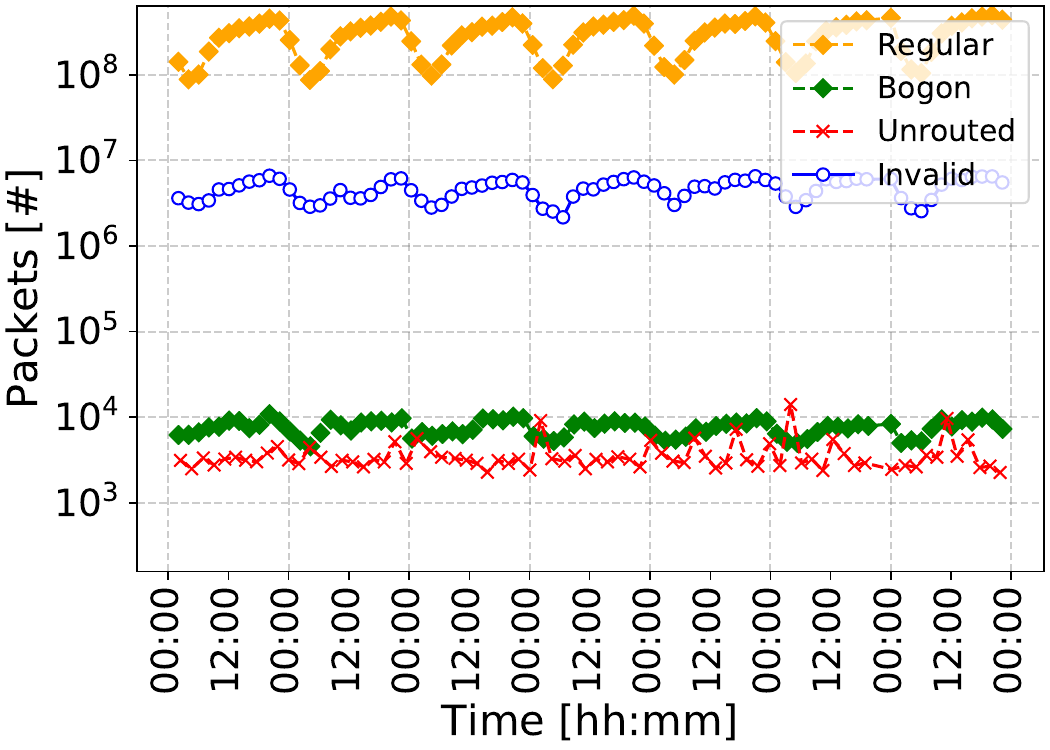}
    \caption{CAIDA Customer Cone}
    \label{fig::week_cc}
  \end{subfigure}
  
  \bigskip
  \begin{subfigure}{.48\textwidth}
    \includegraphics[width=\linewidth]{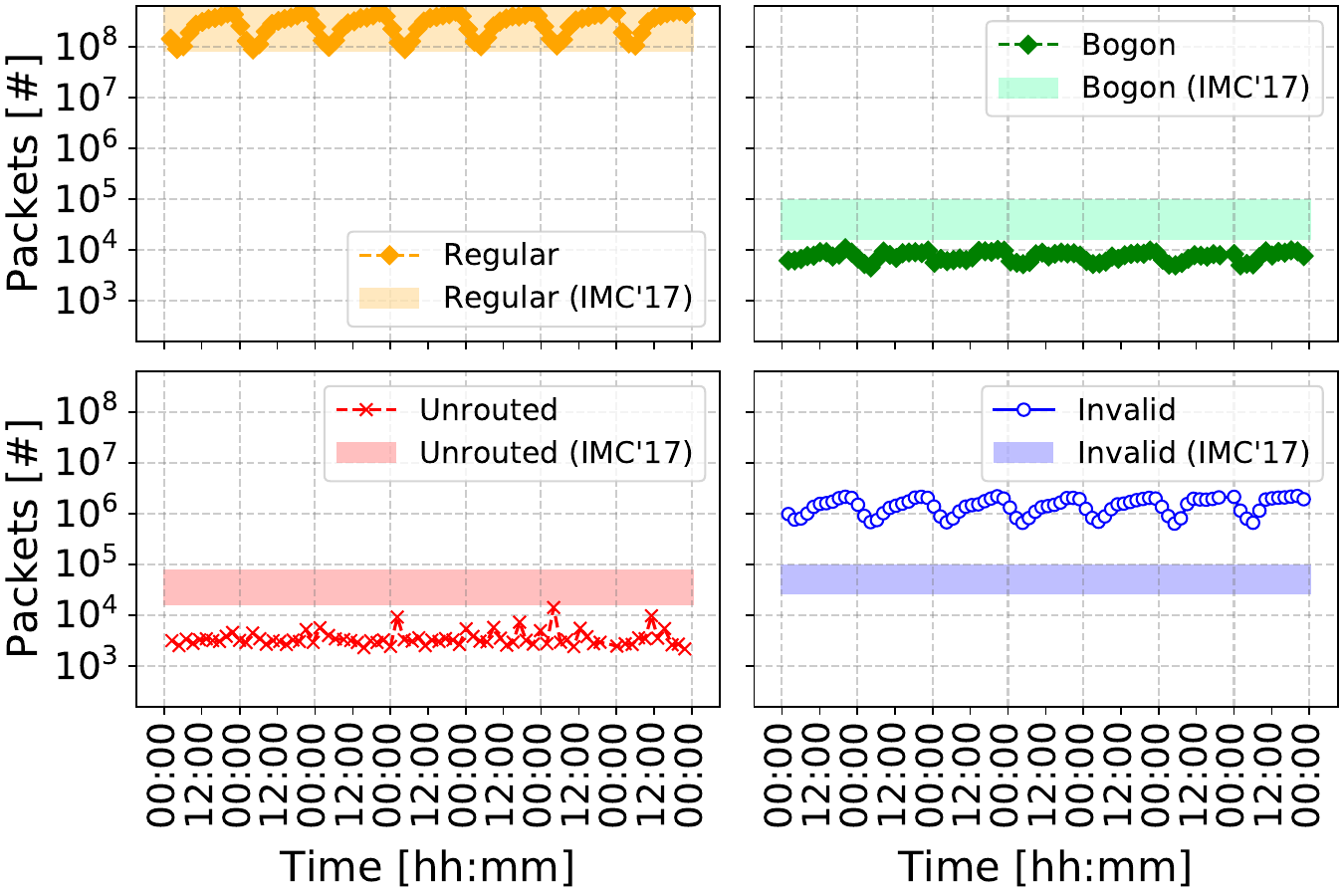}
    \caption{Full Cone}
    \label{fig::week_full}
  \end{subfigure}\hfill
  \caption{Time series of classified traffic distributions}
  \label{fig::week}
\end{figure}

\begin{figure*}
  \begin{subfigure}{.33\textwidth}
    \centering
    \includegraphics[width=\linewidth]{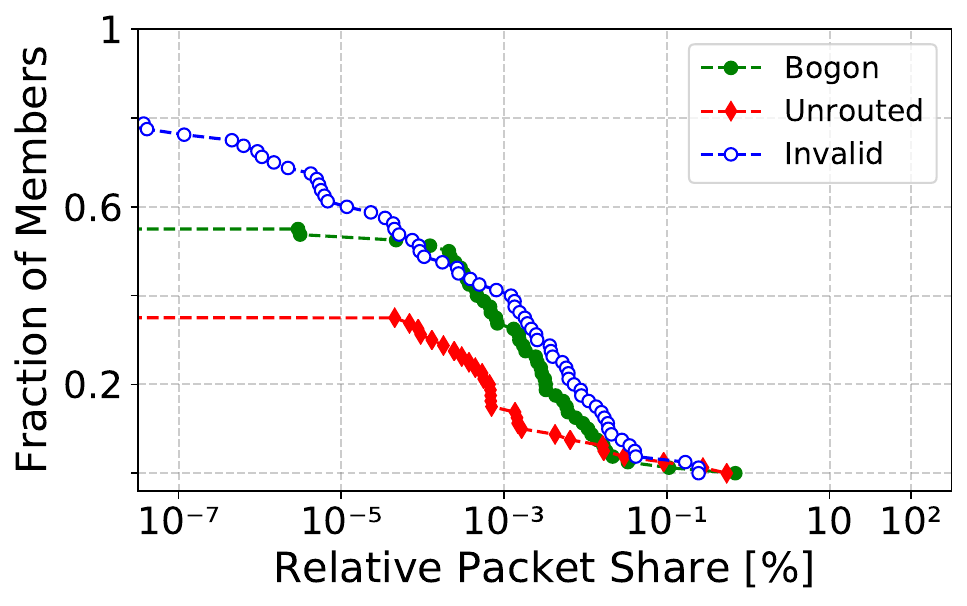}
    \caption{Naive Approach}
    \label{fig::traffic_naive}
  \end{subfigure}\hfill
  \begin{subfigure}{.33\textwidth}
    \centering
    \includegraphics[width=\linewidth]{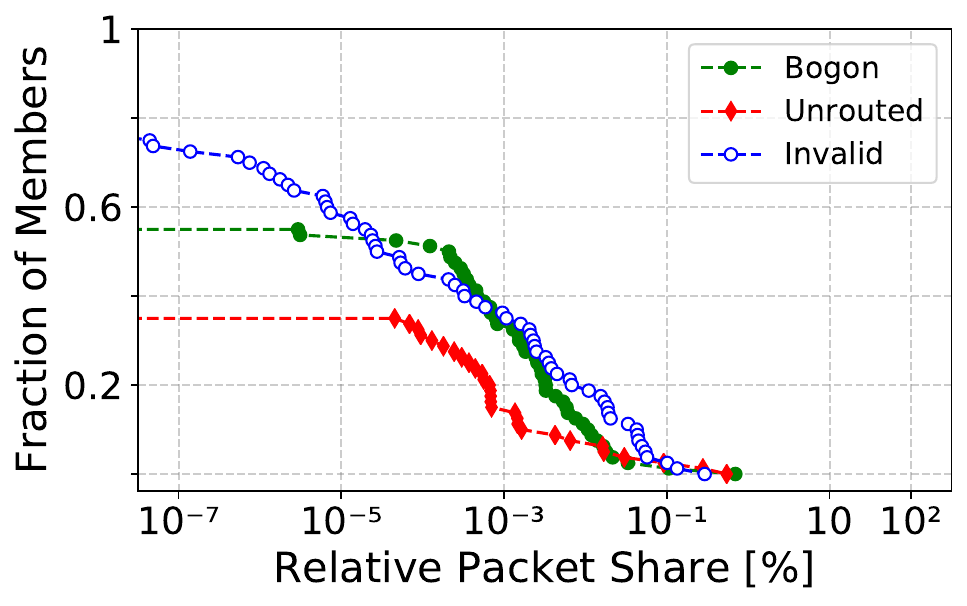}
    \caption{CAIDA Customer Cone}
    \label{fig::traffic_cc}
  \end{subfigure}\hfill
  \begin{subfigure}{.33\textwidth}
    \centering
    \includegraphics[width=\linewidth]{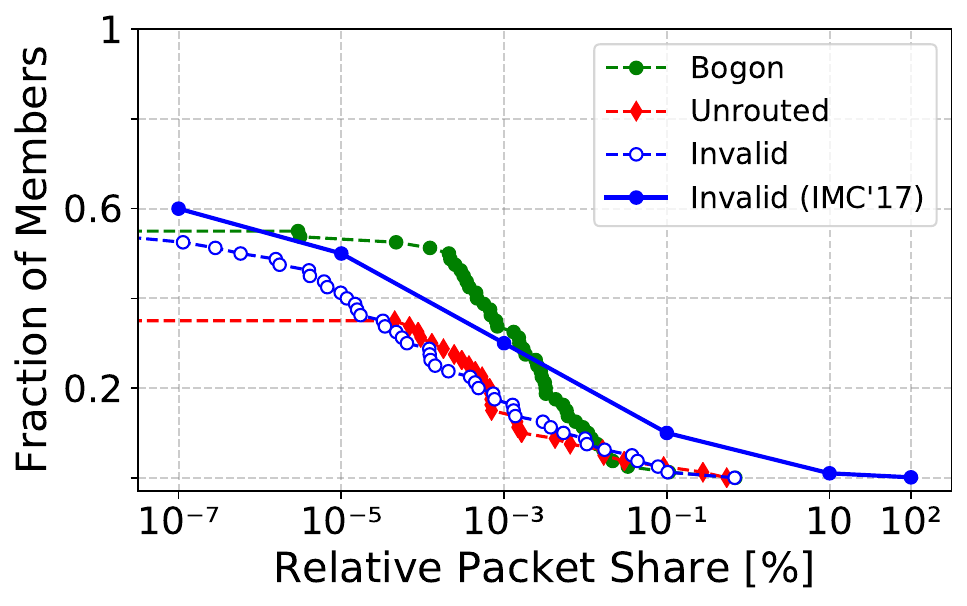}
    \caption{Full Cone}
    \label{fig::traffic_full}
  \end{subfigure}\hfill
  \caption{CCDF of the fractions of anomalous traffic per IXP member AS}
  \label{fig::traffic}
\end{figure*}

\begin{figure*}
  \begin{subfigure}[b]{.33\textwidth}
    \centering
    \includegraphics[width=\linewidth]{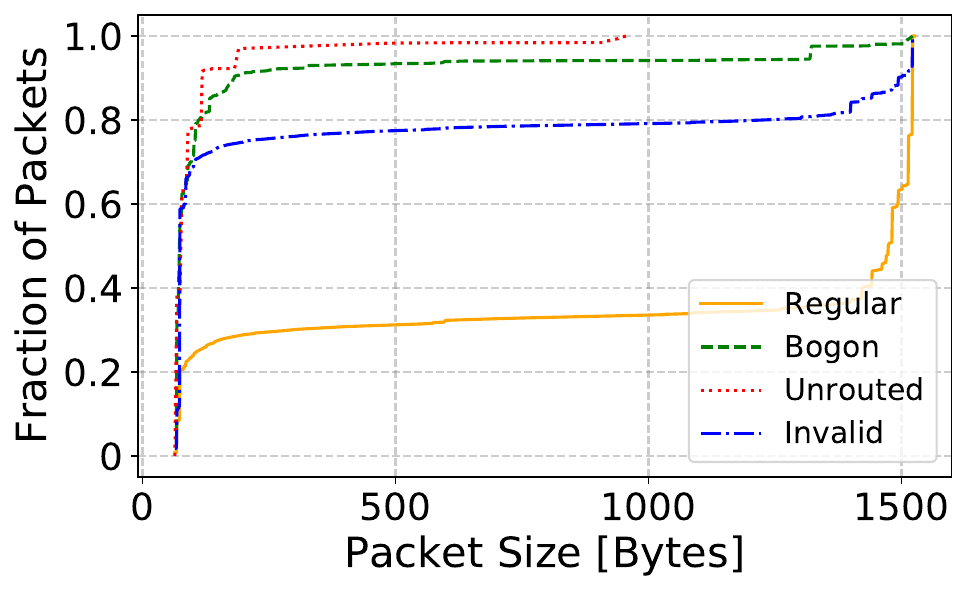}
    \caption{Naive Approach}
    \label{fig::sizes_naive}
  \end{subfigure}\hfill
  \begin{subfigure}[b]{.33\textwidth}
    \centering
    \includegraphics[width=\linewidth]{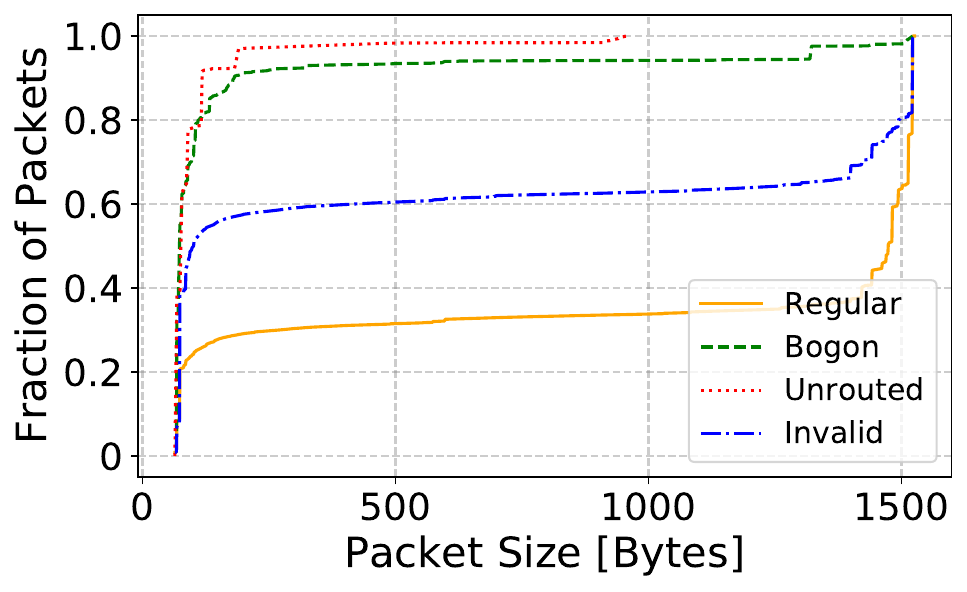}
    \caption{CAIDA Customer Cone}
    \label{fig::sizes_cc}
  \end{subfigure}\hfill
  \begin{subfigure}[b]{.33\textwidth}
    \centering
    \includegraphics[width=\linewidth]{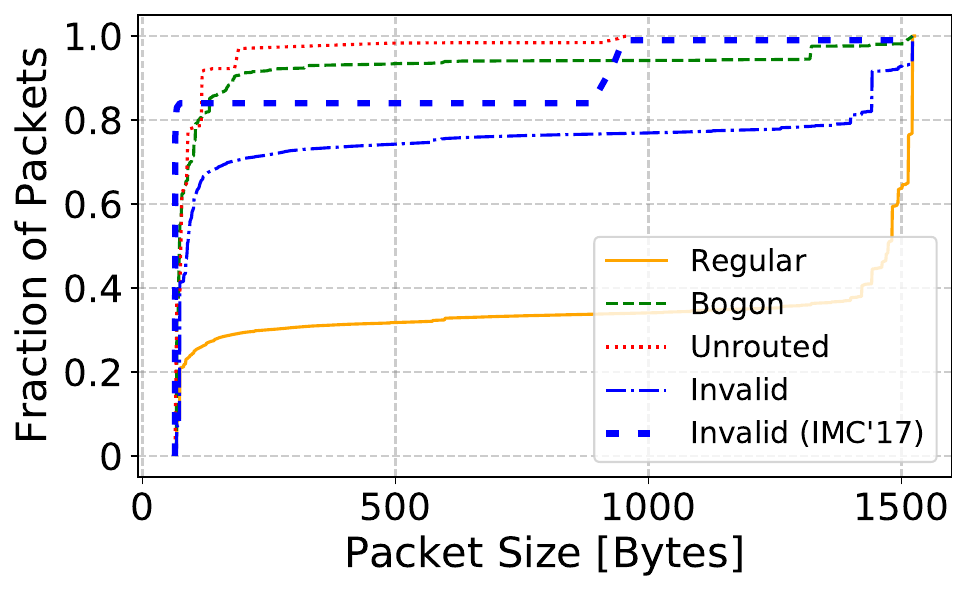}
    \caption{Full Cone}
    \label{fig::sizes_full}
  \end{subfigure}\hfill
  \caption{CDF of packets sizes by category}
  \label{fig::sizes}
\end{figure*}

\subsection{Discrepancies for Invalid Traffic}

Figure~\ref{fig::week} displays the underlying time series of classified traffic for the different cone types. Absolute numbers have been rescaled by a common factor to fit the IMC'17 range. Corresponding results from IMC'17, which are only available for the full cone approach, are indicated by the shaded regions.   

It is clearly visible that the different, stepwise increasing of cones enfold a moderate impact on the traffic fraction that is classified as invalid. Packet numbers marked as invalid drop less than one order of magnitude when moving from the naive to the full cone approach. This is in contrast to the IMC'17 results, which show invalids of almost two orders of magnitude less than our results for the full cone.  

We compare the fractions of anomalous traffic as contributed from the IXP members in Figure~\ref{fig::traffic}. A large group of IXP members issue only a very small portion of invalid packets---some of which disappearing for the full cone approach. Still no  member has more than 1\% of its traffic classified as either bogon, unrouted, or invalid. This is in strong contrast to the results of IMC'17, where a few members emit close to 100 \% of their entire traffic to invalid, i.e., almost all traffic of these ASes is identified as spoofed (cf., Fig. 4 in~\cite{lskrf-dcait-17}).

Given these significant discrepancies between our reproduced results and IMC'17 for the fraction and the distribution of traffic classified as invalid, we question the correctness of classification by taking a closer look at the invalid packets.  

First, we inspect the observed packet sizes per category. Considering that spoofed packets are often used for amplification attacks, a larger amount of small invalid packets  would be in support of the classification results. Figure~\ref{fig::sizes} shows the packet size distributions.

All three approaches exhibit a similar distribution of regular packet sizes with most packets larger than 1200 bytes. In contrast, bogon and unrouted traffic is overwhelmingly made up of small packets. Invalid packets tend to be smaller but vary between the approaches. Still sizes of invalid packets show a wide distribution with significant portions of large packets. This again is in contrast to the IMC'17 results, which show a sharp cut-off for packet sizes larger than $\approx 900$~Bytes.

\subsection{Traffic Mix Reveals False Positives}

\setlength{\tabcolsep}{1.4pt}
\begin{table}
	\caption{Traffic mix per protocol and destination port of invalid packets from the reproduced full cone}
	\label{tab::traffic_mix}
	\begin{tabular}{l c c c c c c c c}
		\toprule
		 \multirow{2}{*}{ICMP}
		 &             &        &    &     &  &  &      & total  \\
                &  &   &   &  &   &   &   &  0.37~\% \\\midrule
            
		\multirow{2}{*}{UDP} 
			&     53   &    123  & 161     &  443 &     19302 &    eph. &  other & total \\
		    &  1.18~\% & $< 0.1$~\% & 0.35~\% & 19.73~\% &  0.18~\% &   0.94~\%  & 0.81~\% & 20.36~\%\\\midrule

		\multirow{2}{*}{TCP}
		& 80 &  443     &       27015 &     10100 &   & eph. &  other & total  \\
                & 3.50~\% &  62.29~\% & 0.00~\% &  0.00~\% &  -- &  6.75~\% & 13.67~\% & 79.45~\% \\

		\bottomrule
	\end{tabular}
\end{table}
\setlength{\tabcolsep}{4pt}

We are now diving deeper into packet inspection of the traffic classified as invalid and want to understand its characteristics.    Table~\ref{tab::traffic_mix} explores the traffic mix and lists the top destination port distributions of invalid UDP and TCP packets. We cannot equivalently compare to the IMC'17 results, as their traffic mix has not fully been disclosed.

Strikingly, we find the majority of invalid traffic to be HTTPS over TCP followed by Quick (using UDP), and plain HTTP according to the transport ports. Typical amplification/reflection attack patterns such as DNS (UDP 53) and NTP (UDP 123) do not stand out in our data, even though they were reported to dominate in the IMC'17 results. On the overall, almost 80~\% of TCP traffic appears bidirectional (\eg is encrypted or carries data) and hence raises doubts about its spoofed nature. Rather, this strongly indicates that the traffic classified as invalid from our data set  mainly consists of regular Web flows and hence has been  classified  erroneously based on the previously proposed methodology.

\begin{table*}[htb!]
  \caption{False positive indicators in traffic of the reproduced full cone}
  \label{tab::false_positve_idicators}
  \begin{tabular}{ l r r r r r r}
    \toprule
                          & SSL over TCP              & HTTP response             & ICMP echo reply           & TCP ACK                    & malformed \\
    \midrule
    Naive Approach        &                  3.985\%  & \cellcolor{Gray}{0.174\%} &                  0.056\%  & \cellcolor{Gray}{86.188\%} &   0.000\% \\
    CAIDA Customer Cone   &                  4.166\%  &                  0.134\%  &                  0.070\%  &                  69.197\%  &   0.000\% \\
    CAIDA (multi-AS ext.) &                  4.166\%  &                  0.134\%  & \cellcolor{Gray}{0.081\%} &                  80.148\%  &   0.000\% \\
    Full  Cone            &  \cellcolor{Gray}{6.395\%}  &                  0.117\%  &                  0.043\%  &                  76.079\%  &   0.001\% \\
    Full (multi-AS ext.)  & \cellcolor{Gray}{6.512\%} &                  0.029\%  &                  0.044\%  &                  77.350\%  &   0.001\% \\
	  Full (w/ Spoofer-IX) & 	      5.776\% &  \cellcolor{Gray}{0.187\%}&		     0.052\%  & 	68.507\%
 &   0.000\% \\    \bottomrule
  \end{tabular}
\end{table*}

\subsection{Comparing with Spoofer-IX}\label{sec:spoofer}

Recent, independent work by Müller et al.~\cite{mlhcb-cisti-19} took the attempt to overcome shortcomings of the IMC'17 work by including AS relationships in modeling the full cone. They  correctly observe that provider-to-customer traffic is always regular, not bound to any customer cone, and hence unverifiable. A measurement study performed at a regional IXP in Brazil shows that among all traffic at the IXP a significant fraction withstands source address validation. Unverifiable traffic in~\cite{mlhcb-cisti-19} indeed appears at the same order of magnitude as regular traffic.

\begin{figure}
    \includegraphics[width=\linewidth]{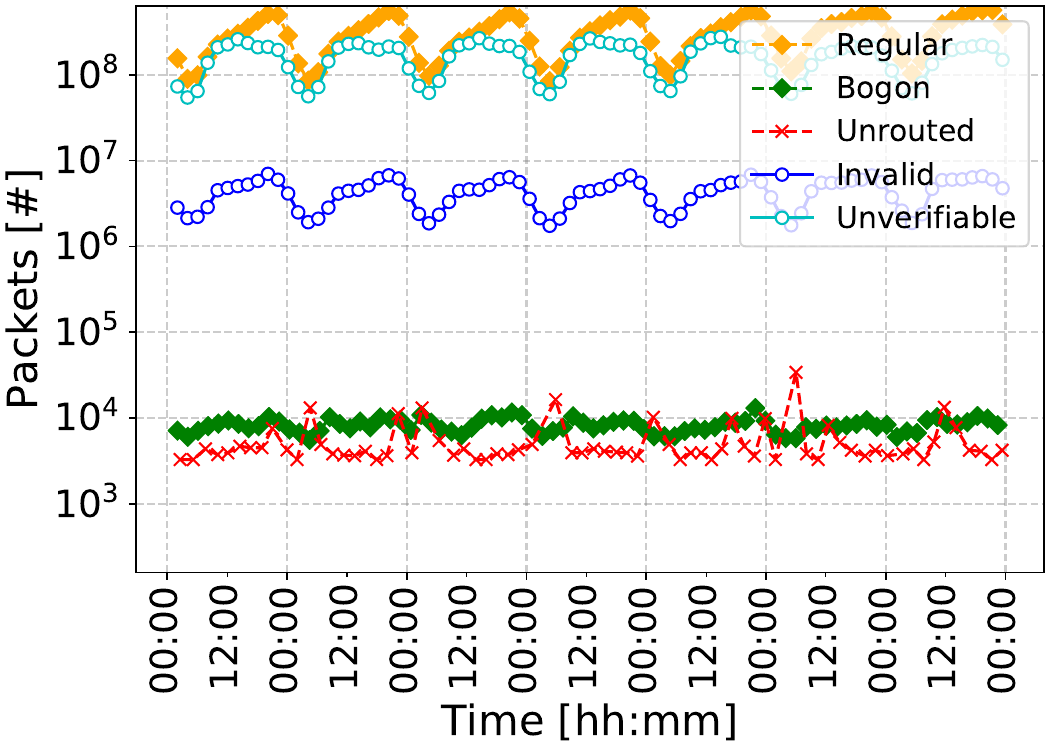}
  \caption{Time series of traffic classification including unverifiable as derived from Spoofer-IX}
  \label{fig::week_spoofer}
\end{figure}

We see similar traffic behavior in our attempts to reproduce the Spoofer-IX results. We perform this step to answer the question whether accounting for the peering relations as mandated by Spoofer-IX can actually repair the methodology proposed in IMC'17. Our results for the continuous traffic classification are shown in Figure~\ref{fig::week_spoofer}. Clearly visible is  a large fraction of unverifiable traffic---about one third of the regular. Nevertheless, the amount of traffic identified as invalid (spoofed) remains very similar to our previous results from using the CAIDA customer cone.

To further  assess the invalid traffic as classified from the different approaches, we test for specific indicators of unspoofed traffic. We selected the five indicators (1) SSL over TCP, (2) HTTP responses, (3) ICMP echo replies, (4) TCP packets carrying ACKs, and (5) malformed packets (\eg transport port 0, same SRC and DST address) as used by CAIDA~\cite{dbkck-eiasu-14}. These indicators are not rigorous, but strong.  
Injecting packets into existing TCP connections requires some guesswork and is not easily deployed on a large scale. TCP packets that carry HTTP responses or ACKs might be less likely to be spoofed. The presence of an encrypted channel only strengthens this assumption. Packets that can easily be dropped by the receiver and neither provoke a reply nor require action are not attractive for reflection attacks either. ICMP echo replies are an example for this category. Looking at the problem from the opposite direction, malformed packets that could disrupt communication are more likely to be spoofed than part of regular traffic. As part of this analysis, we looked for packets that use port zero or the same destination and source address.

Our findings are summarized in Table~\ref{tab::false_positve_idicators}. They clearly indicate that for any selected cone more than 75~\% of the packets carry a distinct characteristic of unspoofed traffic. We conclude that applying the algorithmic method of spoofing detection at IXPs  presented in \cite{lskrf-dcait-17} cannot be reproduced on a quantitative scale, and most likely leads to results dominated by false positives. We also conclude that the appropriate separation of provider-to-customer traffic as devised by Spoofer-IX remains fairly effectless.

\section{Discussion}\label{sec:discussion}

An algorithm that can automatically detect and eliminate spoofed packets at the Internet core, \ie at an IXP, gives hope to Internet operators to overcome one of the major security threats. While significant efforts, costs, and collateral damages~\cite{hnjds-pfdms-18b,nbdsw-dbhdo-19} are accepted today to defend against DDoS attacks, these drawbacks could be largely avoided by simply deploying such spoofing detectors at central exchange points. Unfortunately, the results of this reproducibility study do not support the promise to solve this problem.

We investigated the reasons for these disappointing results. Pursuing additional analyses and private communication with the authors, we could identify a number of unresolved obstacles.  Aside from minor inaccuracies of the IMC'17 method~\cite{ietf-private-19} such as \one inaccurate timing with respect to control and data plane measurements, \two  disregarding of BGP withdraw messages in the public dump files, and \three inaccurate modeling of transitivity in the BGP routing graph~\cite{swscb-cuflsr-16}, the construction of the customer cones always suffers irremediably  from the lack of visibility of BGP relations. 

Transferring these general shortcomings to the specific observation perspective of an IXP reveals additional structural problems. 
An IXP provides layer~2 cross-connecting infrastructure that remains transparent to the routing layer. An IXP takes no role in the routing topology, and thus cannot sensibly be used as a reference point for control plane analyses. Virtually any kind of inter-domain traffic traverses the IXP switch fabric, including that of pure transit, of  traditional business relations~\cite{g-iasri-01}, of newly emerging business relations (\eg paid peering), agreements of regional communities, and private peerings. 

Previous work~\cite{kkz-didmh-13} already excluded transit traffic from source address validation. Independent, parallel work~\cite{mlhcb-cisti-19} correctly coined provider-to-customer traffic `unverifiable' as discussed in Section~\ref{sec:spoofer}. Authors, however, could not sharpen the filters for source address validation at IXPs, because most customer cones of providers (\eg Tier-1, NSPs) are too large to sensitively filter out spoofers. Very recent work of Fonseca et al.~\cite{fcfmj-tdssi-20} also observed that most customer cones present at IXPs are too large to be of use. Instead, authors propose to identify origins of spoofers by varying anycast announcements for the detectors in BGP and infer source networks of the arriving traffic location. 

In this paper, we want to emphasize that traffic governed by (hidden) private contracts or bilateral agreements may cross the IXP platform,  of which control information never enter public announcements. 
One illustrative example (observed in the wild) is the following: Provider A and its customer B are both physically present at an IXP. B obtains provider aggregated (PA) address space from A and consequently should not announce these prefix(es) independently in public. Still B can legitimately forward packets with source addresses from the PA space directly across the IXP. Hence, B with its delegated PA space does not appear in any customer cone, why all packets originating from B with these source addresses  traversing the IXP (to A or a private peer) are incorrectly classified as invalid (spoofed) by the IMC'17 method. This sample case explains the `anomalies' of ASs that emit 100\% spoofed packets as observed in IMC'17. It is noteworthy that RPKI deployment would require issuance of de-aggregated IP prefixes in ROAs~\cite{RFC-6482} for any additional peering of B.

The authors of IMC'17 seem to have been unaware of such intricacies present at common IXP platforms. They chose to mitigate these problems by manually inspecting the traffic flows and adjusting peering relations within the customer cones accordingly (\cf Section~4.4, ``Missing AS Links:'' in~\cite{lskrf-dcait-17}). Such manual adjustments of data sets are understandable for the individual case under exploration. They discard, however, the presented method from future use in a real-world deployment.

\section{Conclusion}
\label{sec:conclusion}

In the present study, we aimed at reproducing the results of methods for identifying spoofed traffic at IXPs that were presented at IMC'17. Using different but equivalent data sets, scripts provided by the authors, as well as an extended, independently written tool set, we could reproduce certain general observations of the paper and pursue an extended, thorough analysis of the various results.


Following a purely algorithmic approach, however, we could not reproduce the results of IMC'17. Instead, the majority of packets identified as ``invalid'' (spoofed) appeared as false positives. Our further analysis of the problem space and the specific outcomes on the control and the data plane revealed a structural shortcoming of the proposed method. Source address validation of data packets against the control plane is more intricate than considered.
  
In trying to reproduce the paper \textit{Detection, Classification, and Analysis of Inter-Domain Traffic with Spoofed Source IP Addresses}~\cite{lskrf-dcait-17}, we found that the manual component of the methodology \one represents the major challenge in terms of reproducibility, and has \two a significant effect on the results; this highlights challenges in deploying approaches based on current methods in an automated~fashion.

\subsection*{Acknowledgments}
We explicitly thank Franziska Lichtblau and Florian Streibelt for patiently answering our questions during this  study.
Philipp Richter is gratefully acknowledged for concluding remarks.
We thank Alberto Dainotti and Alistair King for organizing the DUST workshop at CAIDA, which paved the ground for discussing the results in a wider scope.
Finally, we thank the reviewers of the ACM IMC 2019 Poster Track for their valuable feedback.

This work was supported in parts by the German Federal Ministry of Education and Research (BMBF) within the project \emph{X-Check}.

\balance

\bibliographystyle{ACM-Reference-Format}
\bibliography{rfcs,security,internet,own,local,tmp}

\end{document}